\documentclass[aps,reprint,longbibliography,superscriptaddress]{revtex4-1}
\usepackage{amsmath}
\usepackage{amssymb}
\usepackage{bm}
\usepackage{epsfig}
\usepackage{graphicx}
\usepackage[dvipsnames]{xcolor}
\usepackage{color}
\usepackage[unicode=true,colorlinks=true,citecolor=blue,urlcolor=blue]{hyperref}
\usepackage[T2A]{fontenc}
\usepackage[cp1251]{inputenc}
\usepackage[english]{babel}
\usepackage{braket}
\usepackage{physics}

\graphicspath{{img/}}

\let\ifr\i

\renewcommand{\Re}{\mathop{\rm Re}}
\renewcommand{\Im}{\mathop{\rm Im}}

\newcommand{\e}{\mathrm{e}}

\renewcommand{\i}{{\rm i}}
\renewcommand{\d}{\mathrm d}
\renewcommand{\emph}{\textit}

\renewcommand{\braket}[1]{\left\langle #1 \right\rangle}

\begin{document}

\title{Quantum Zeno effect under continuous spin noise measurement\\ in a quantum dot-micropillar cavity}

\author{N.~V.~Leppenen}
\affiliation{Ioffe Institute, 194021 St. Petersburg, Russia}

\author{L.~Lanco}
\affiliation{Centre for Nanosciences and Nanotechnology, CNRS, Universite Paris-Saclay, UMR 9001, 10 Boulevard Thomas Gobert, 91120, Palaiseau, France}
\affiliation{Universite Paris Diderot - Paris 7, 75205 Paris CEDEX 13, France}

\author{D.~S.~Smirnov}
\email{smirnov@mail.ioffe.ru}
\affiliation{Ioffe Institute, 194021 St. Petersburg, Russia}


\begin{abstract}
  We theoretically describe the quantum Zeno effect in a spin-photon interface represented by a charged quantum dot in a micropillar cavity. The electron spin in this system entangles with the polarization of the transmitted photons, and their continuous detection leads to the slowing of the electron spin precession in external magnetic field and induces the spin relaxation. We obtain a microscopic expression for the spin measurement rate and calculate the second and fourth order correlation functions of the spin noise, which evidence the change of the spin statistics due to the quantum Zeno effect. 
We demonstrate, that the quantum limit for the spin measurement can be reached for any probe frequency using the homodyne nondemolition spin measurement, which maximizes the rate of the quantum information gain.
\end{abstract}
\maketitle

\section{Introduction}

The quantum Zeno effect in the most spectacular way shows the fundamental difference between the quantum microscopic world and the classical everyday life. Explicitly, it stands that the continuously observed quantum object can not move~\cite{khalfin1958contribution,Facchi_2008}. After being formulated as a paradox for the classical objects by Zeno of Elea in the 5th century BC~\cite{Aristotle}, it was later described as a physical effect for the quantum objects in the most popular way in 1977~\cite{doi:10.1063/1.523304}.

Nowadays many aspects of the quantum Zeno effect are under active investigation~\cite{RevModPhys.91.045001,RevModPhys.90.035005}. These include, for example, deceleration of the quantum dynamics under continuous weak measurement~\cite{Gross_2018}, quantum anti-Zeno effect~\cite{Kofman2000,PhysRevLett.86.2699}, dynamics in the quantum Zeno subspaces~\cite{RevModPhys.90.031002}, and observation of the quantum Zeno effect in various systems from free atoms to semiconductors~\cite{RevModPhys.75.281,physreva.88.020101,kalb2016experimental,pfender2019high,cujia2019tracking}. The general interest is additionally boosted by the importance of this effect for the quantum computation~\cite{Hosten2006}. The quantum Zeno effect on one hand can help to increase the storage time of the quantum information~\cite{PhysRevLett.108.080501}, but on the other hand it can slow down or even damage the computations~\cite{stolze2005quantum}.

Despite the fundamental importance of the quantum Zeno effect, most of its descriptions have a very general form and their validity to every specific system is always questionable~\cite{PETROSKY1990109}. Generally, the quantum Zeno effect stems from the interaction of the system with the environment. But in experimental observation of an effect it is not always clear if the interaction with the environment really represents the measurement that yields the quantum information or not~\cite{Clerk}.

In this work we consider a spin photon interface, which is a basic building block for quantum communication, simulations and cryptography~\cite{knill2001scheme,Loredo2019}. It is represented by a quantum dot (QD) micropillar cavity with a resident electron in the QD. The spin dynamics in this system can be conveniently studied using the spin noise spectroscopy~\cite{Zapasskii:13}. In this recently developed technique, the linearly polarized light is incident at the device and the circular polarization degree of the transmitted light is continuously measured. It is proportional to the instantaneous electron spin polarization. The Fourier transform of the autocorrelation function of the light polarization degree is proportional to the spin noise spectrum of the electron~\cite{BackAction}. This technique allows one to access the spin properties such as $g$ factor and spin relaxation time without system excitation.
%

Recently it was suggested that with increase of the spin measurement strength (intensity of the probe light), the spin dynamics undergo a quantum dynamical phase transition due to the quantum Zeno effect~\cite{doi:10.1063/1.2193518}. The transition can be evidenced in the modification of the counting statistics, which is described by the generating functions~\cite{Li_2014}. They satisfy the Schr\"{o}dinger-like equation with the non-Hermitian Hamiltonian. The eigenspectrum of this Hamiltonian can be described by a certain braid group~\cite{PhysRevE.87.050101}. When the quantum Zeno effect takes place, the spin precession in the magnetic field gets suppressed~\cite{noise-trions}, and the braid group of the eigenspectrum changes. So the spin dynamics arguably undergoes the topological phase transition.


In our work we theoretically demonstrate the quantum Zeno effect in a QD micropillar cavity under continuous interaction of spin with light.
We show that one can use the phenomenological approach to its description in the limit of weak magnetic field and small population of the excited states. We explicitly express the spin measurement strength through the microscopic parameters of the device such as the intensity of the light, trion optical transition frequency, and the light matter coupling strength. Further we calculate the quantum information gain rate and find that it is determined by the concurrence of electron and photon spins and it is of the same order as the measurement strength. We also demonstrate that homodyne detection allows one to reach the quantum limit for the nondemolition spin measurement, when the measurement rate equals to the spin dephasing rate. Finally, we study the modification of the spin statistics under continuous weak measurement and demonstrate that it qualitatively changes at the point of the transition between the phases of spin oscillations and monotonous spin decay.

Our paper is organized as follows. In the next section we formulate the model of the device under study. Then in Sec.~\ref{sec:results} we describe the quantum Zeno effect using the three independent approaches: phenomenological, numerical and analytical, and establish the relation between them. In the following Sec.~\ref{sec:informatics} we calculate the quantum nondemolition spin  measurement rate, relate it with the entanglement between electron and photons spins, and compare with the spin dephasing rate. We also find the conditions, when the quantum limit for the spin measurement is reached. Finally, in Sec.~\ref{sec:statistics} we study the statistics of the spin noise and calculate the noise bispectrum. 
Applicability of our findings to the realistic devices and the brief summary are presented in the concluding section~\ref{sec:concl}.

\section{Model}
\label{sec:model}

\begin{figure}[t]
  \centering
  \includegraphics[width=1\linewidth]{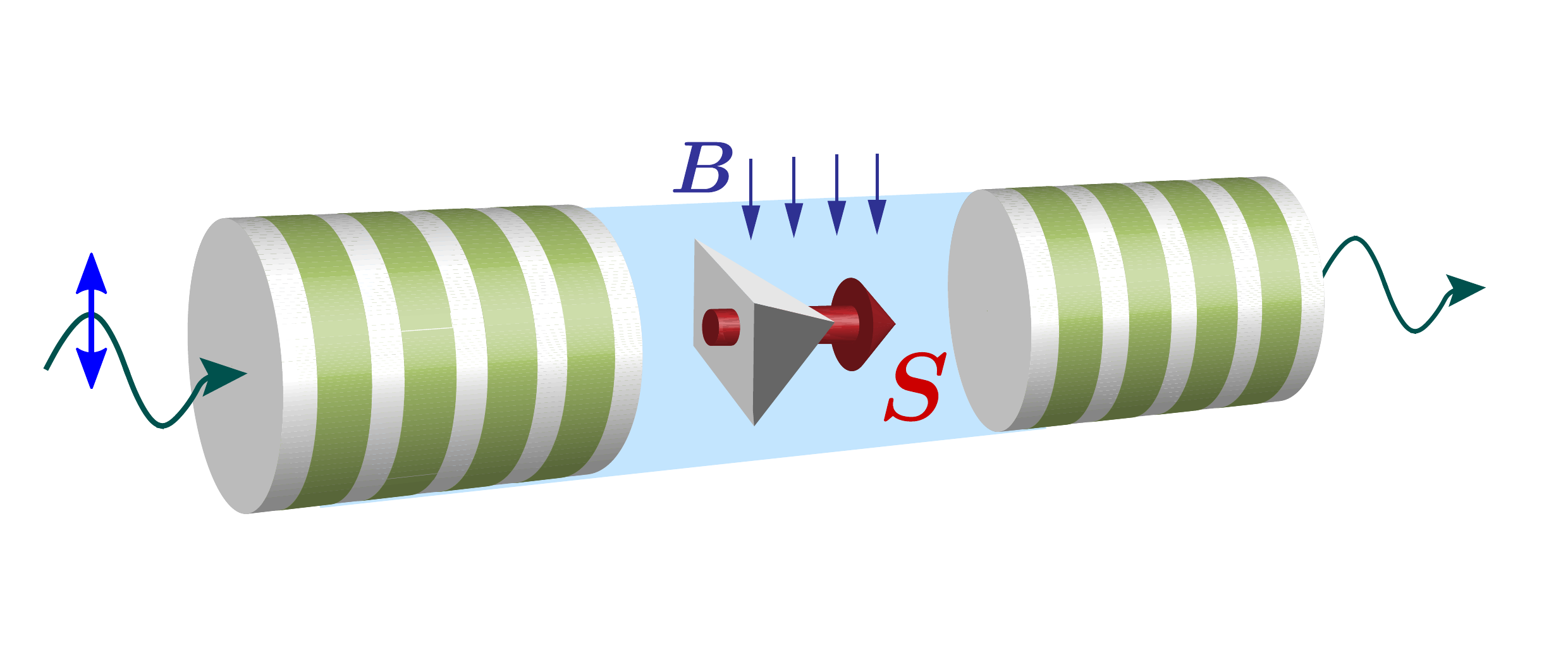}
  \caption{Sketch of a QD micropillar cavity with a single electron with the spin $\bm S$ in external magnetic field $\bm B$ with the incident linearly polarized probe light.}
  \label{fig:sketch}
\end{figure}

We consider a micropillar cavity with a QD inside it, Fig.~\ref{fig:sketch}. We assume that the QD is charged with a single electron. The device is placed in an external transverse magnetic field (perpendicular to the structure growth axis $z$) and coherently excited by the continuous linearly polarized light.

The Hamiltonian of the system has the form~\cite{singleSpin,BackAction}:
\begin{equation}
 \label{ham}
 {\cal H} = {\cal H}_{+}+{\cal H}_{-}+{\cal H}_{B}.
\end{equation}
The first two terms describe the contributions with right-handed and left-handed helicities (signs of the angular momentum projections on the $z$ axis): 
\begin{multline}
  \label{eq:H_pm}
	{\cal H}_{\pm} =\hbar\omega_c c_\pm^\dag c_\pm + \hbar\omega_0 a_{\pm 3/2}^\dag a_{\pm 3/2}+\\+
 \hbar\left(gc_\pm a_{\pm 3/2}^\dag a_{\pm 1/2}
+ \mathcal{E}_\pm e^{-i\omega t} c_\pm^\dag + {\rm H.c.} \right).
\end{multline}
Here $c_\pm$ ($c_\pm^\dag$) are the annihilation (creation) operators of the $\sigma^\pm$ photons in the cavity. The two orthogonally polarized cavity modes have the eigenfrequency $\omega_c$, they are assumed to be degenerate. The QD is described by the four states with the corresponding annihilation operators $a_{\pm 1/2}$ and $a_{\pm 3/2}$. The former two correspond to the two ground electron states with the spin projection $S_z=\pm1/2$. The latter two correspond to the excited singlet trion states with the energy $\hbar\omega_0$. The trion consists of two electrons with the opposite spins and a heavy hole with the spin $J_z=\pm3/2$. Further, $g$ is the light matter coupling strength. It describes the photon absorption from the cavity mode and the creation of a trion from a single electron, as well as the reverse process. According to the optical selection rules the total angular momentum component along the $z$ axis is conserved, so absorption of $\sigma^\pm$ photon is accompanied by the creation of a hole with $J_z=\pm3/2$ and an electron with $S_z=\mp1/2$, respectively~\cite{ivchenko05a}. Due to the Pauli exclusion principle this is possible only when the resident electron in the QD has $S_z=\pm1/2$, as described by Eq.~\eqref{eq:H_pm}. Finally, the parameters $\mathcal E_\pm$ are proportional to the amplitudes of the $\sigma^\pm$ polarized components of the coherent incident light with the frequency $\omega$~\cite{milburn}. We consider the system excitation by linearly polarized light, which corresponds to $\mathcal E_+=\mathcal E_-=\mathcal E$.

The effect of the external magnetic field is described by
\begin{equation}
  {\cal H}_{B} = \frac{\hbar\Omega_L}{2}\sum_\pm a^\dag_{\pm1/2}a_{\mp1/2},
\end{equation}
where $\Omega_{L}$ is the Larmor frequency of the electron. The transverse Land\'e factor of the heavy hole is very small~\cite{Mar99}, so we neglect it.

The system under study is open, and it should be described using the density matrix formalism. The density matrix $\rho(t)$ satisfies the master equation
\begin{equation}
\label{eq:density_m}
  \dot\rho(t)={i\over \hbar}[\rho(t),\mathcal H]-\mathcal L\lbrace\rho(t)\rbrace\:,
\end{equation}
where the dot denotes the time derivative and the Lindblad superoperator $\mathcal L\lbrace\rho(t)\rbrace$ describes the incoherent processes~\cite{milburn}. We take into account only two of them: nonradiative trion decay with the rate $2\gamma$ and the photon escape from the cavity with the rate $2\varkappa$. They are described by
\begin{multline}
  \label{Lindblad}
  {\cal L}\lbrace\rho\rbrace = \sum_{\pm}\left[ \varkappa \left(c_{\pm}^{\dagger}c_{\pm}\rho+\rho c_{\pm}^{\dagger}c_{\pm}-2c_{\pm}\rho c^{\dagger}_{\pm}\right)\right.+\\ +\gamma\left(a_{\pm3/2}^{\dagger}a_{\pm3/2}\rho+\rho a_{\pm3/2}^{\dagger}a_{\pm3/2}-\right. \\ - \left.\left.2a^{\dagger}_{\pm1/2}a_{\pm3/2}\rho a^{\dagger}_{\pm3/2}a_{\pm1/2}\right)\right].
\end{multline}
Note, that $\gamma$ can also account for the radiative trion recombination, if the photon is not emitted into the cavity mode. We assume that the trion recombination conserves the helicity, so after the recombination of a trion with $J_z=\pm3/2$ an electron with $S_z=\pm1/2$ is left in the QD. The cavity mode decay rate is contributed by the photon escape through the left and right mirrors and through the side walls: $\varkappa=\varkappa_1+\varkappa_2+\varkappa_0$, respectively. We assume, that the light is incident at the cavity from the left. The amplitude transmission coefficient through the left (right) mirror is proportional to the square root of $\varkappa_1$ ($\varkappa_2$)~\cite{Carmichael}.

\begin{figure}[t]
  \centering
  \includegraphics[width=1\linewidth]{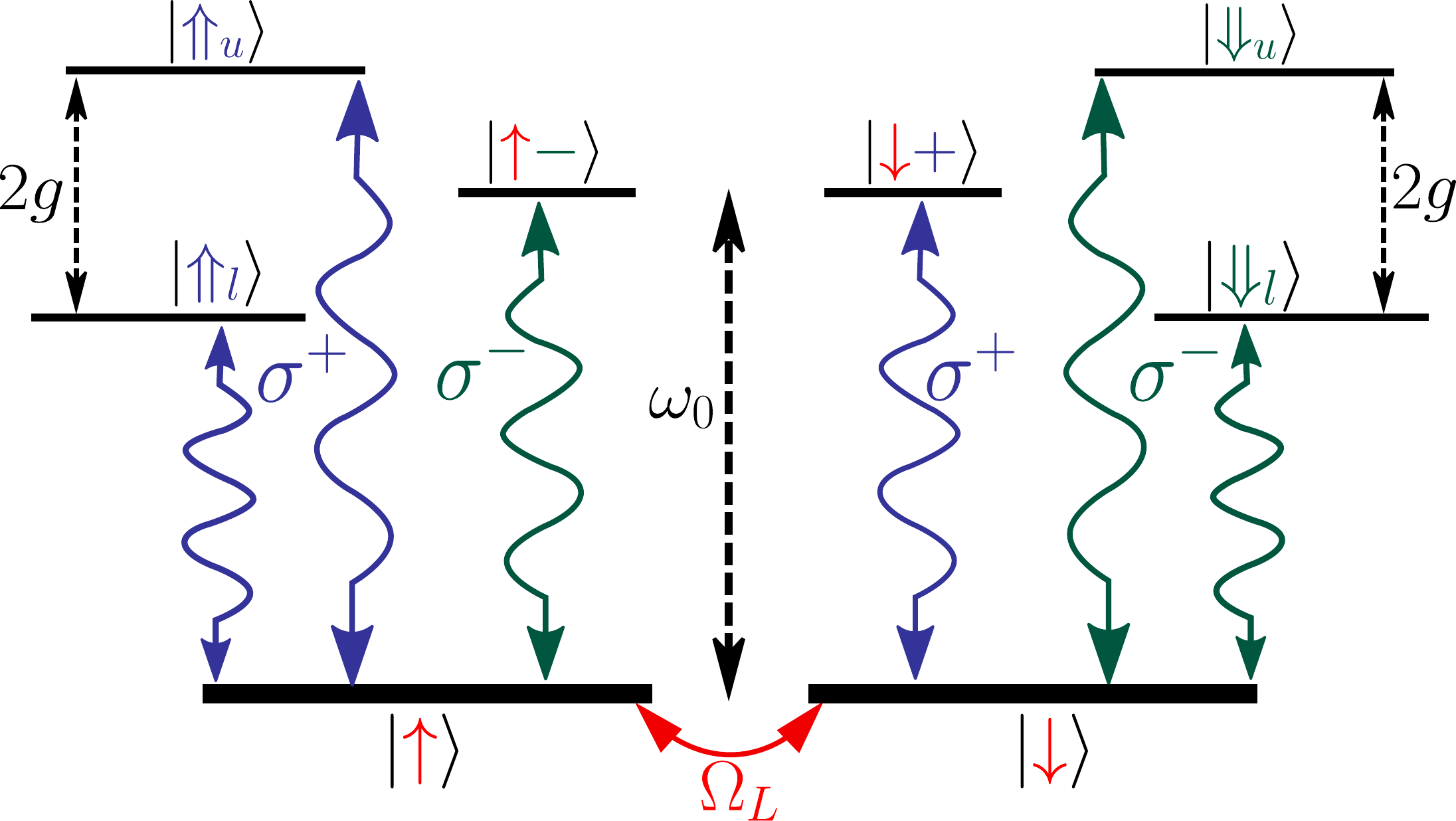}
  \caption{The lowest energy levels of the system and optical transitions between them. The electron spin states are denoted by the red vertical arrows, $\pm$ sign refers to the $\sigma^\pm$ polarization of a single photon, and the polariton states are defined in Eqs.~\eqref{eq:polaritons}.}
  \label{fig:eigen}
\end{figure}

For the rest of the paper we assume the perfect tuning between the trion resonance frequency and the cavity mode: $\omega_c=\omega_0$, and stick to the notation $\omega_0$. Moreover, we limit ourselves to the weak magnetic and driving fields:
\begin{equation}
  \label{eq:weak_field}
  \Omega_L,\mathcal E\ll\varkappa.
\end{equation}
The lowest eigenstates of the system and transitions between them are shown in Fig.~\ref{fig:eigen}. The mostly populated are the two ground spin states, which are mixed by the magnetic field.

The lowest excited states are the single photon states and the polariton states~\cite{microcavities}. If a single $\sigma^\pm$ photon is present in the cavity and the electron is in the spin down/up state, respectively, then the photon can not be absorbed, as described above, and this is an eigenstate. These states are shown in the middle of Fig.~\ref{fig:eigen}. In the opposite case, when the electron and photon have the same helicities, the multiple photon absorption and reemission by the QD lead to the formation of the polariton states. In the case under study ($\omega_c=\omega_0$) they have the form
\begin{equation}
  \label{eq:polaritons}
  \ket{\Uparrow_{^{u}_{l}}} = \frac{a_{+3/2}^\dag \pm c_+^\dag}{\sqrt{2}}\ket{0},
  \quad
  \ket{\Downarrow_{^{u}_{l}}} = \frac{a_{-3/2}^\dag \pm c_-^\dag}{\sqrt{2}}\ket{0},
\end{equation}
where $\ket{0}$ denotes the vacuum state, and the subscripts $u, l$ refer to the upper and lower polariton states, respectively. These states have the energies
\begin{equation}
  \label{eq:E_pol}
  E_{u,l}=\hbar(\omega_{0}\pm g),
\end{equation}
as shown in Fig.~\ref{fig:eigen}. The polariton states~\eqref{eq:polaritons} are the well defined states in the strong coupling regime only, when the damping is smaller, than the splitting: $\gamma,\varkappa\ll g$. In this paper we mainly focus on this regime, but the formalism developed below is valid for the weak coupling regime as well. However, the quantum Zeno effect in the weak coupling regime can be described without taking into account the quantization of the electromagnetic field~\cite{noise-trions}.

\begin{figure}[t]
\centering
 \includegraphics[width=1\linewidth]{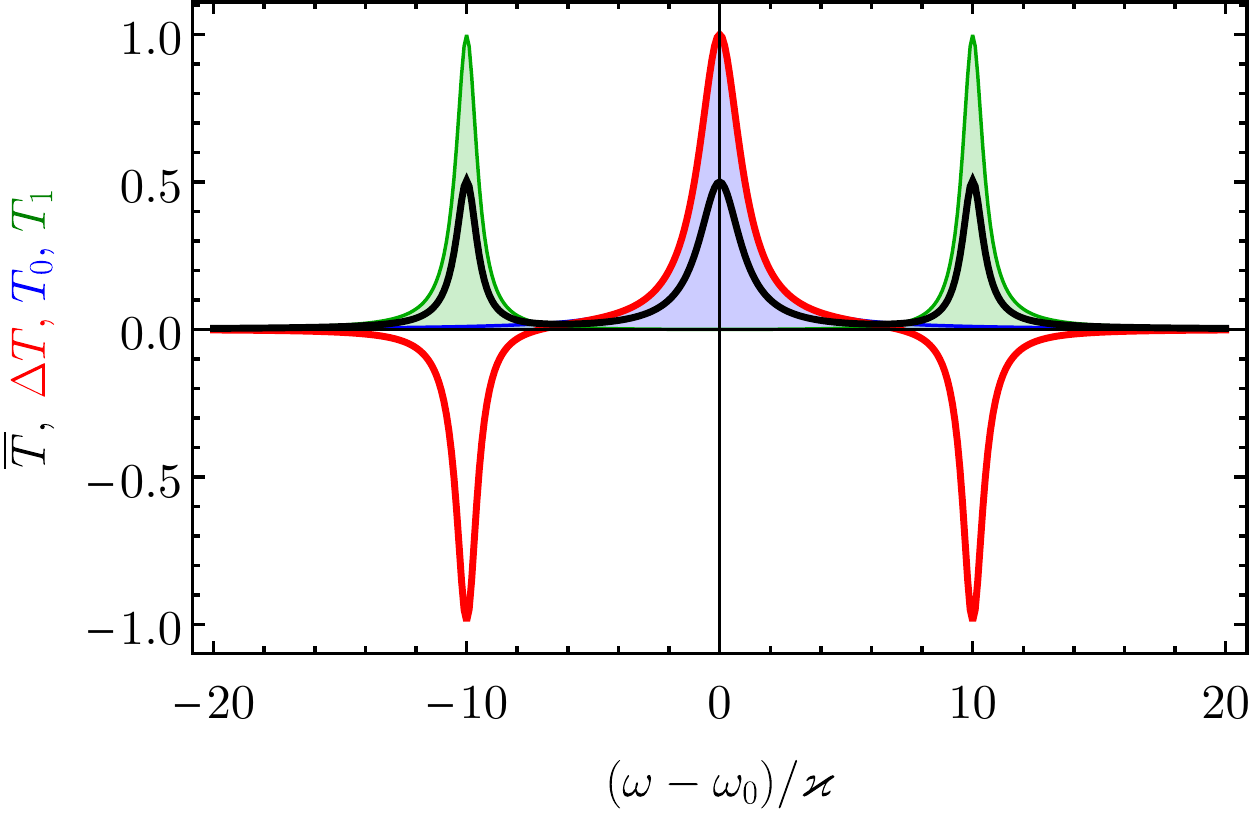}
 \caption{Contributions to the total intensity transmission coefficient $\overline{T}$ and $\Delta T$ [see Eq.~\eqref{eq:T_pm}] (black and red curves, respectively) and the transmission coefficients $T_{0,1}$ (blue and green filled areas, respectively) calculated after Eqs.~\eqref{eq:t01} with the parameters $\gamma=0$ and $g/\varkappa=10$.} 
  \label{fig:trans1}
\end{figure}

For the weak incident light [Eq.~\eqref{eq:weak_field}] the system is mainly in the Hilbert space of the two electron spin states and vacuum photon state. In this case the intensity transmission coefficients of the $\sigma^\pm$ polarized light depend on the electron spin as~\cite{singleSpin}
\begin{equation}
  \label{eq:T_pm}
  T_\pm=\overline{T}\pm \Delta TS_{z}.
\end{equation}
It is convenient to introduce $T_{1,0}=\overline{T}\pm\Delta T/2$, which are the transmission coefficients of circularly polarized light in the case, when the electron has the same/opposite helicity, respectively. They are given by $T_{1,0}=|t_{1,0}^2|$, where~\cite{milburn,PhysRevB.78.085307}
\begin{equation}
  \label{eq:t01}
  t_1=\frac{\i\varkappa}{\omega-\omega_0+\i\varkappa-\dfrac{g^2}{\omega-\omega_0+\i\gamma}},
  \quad
  t_0=\frac{\i\varkappa}{\omega-\omega_0+\i\varkappa}.
\end{equation}
The transmission coefficients are shown in Fig.~\ref{fig:trans1} as functions of the detuning $\omega-\omega_0$ for the strong coupling regime, $g\gg\varkappa,\gamma$. The coefficient $T_0$ is described by a Lorentzian at the bare cavity frequency $\omega_0$. It describes the resonant transmission of the circularly polarized light for the case, when its interaction with the QD is forbidden by the optical selection rules. Another transmission coefficient, $T_1$, describes the resonant light transmission at the polariton energies, Eq.~\eqref{eq:E_pol}, for the case, when the circularly polarized photons can be absorbed by the QD. Typically, the trion decay rate is much smaller than the photon escape rate~\cite{Arnold2015}, so in this figure we consider the limit $\gamma=0$. In this limit, the widths of the peaks in $T_1$ are two times smaller than that in $T_0$, because the polariton consists of the photon only by one half so its decay rate is two times smaller than $\varkappa$.

Eq.~\eqref{eq:T_pm} shows that the transmission coefficient of circularly polarized light depends on the electron spin orientation. The circular polarization degree of the transmitted light for the linearly polarized incident light is proportional to the electron spin polarization:
\begin{equation}
  \label{eq:Sz_I}
  S_z(t)\propto I_+(t)-I_-(t)\equiv\Delta I(t).
\end{equation}
Here $I_\pm(t)$ are the intensities of the transmitted $\sigma^\pm$ polarized light, and we introduced $\Delta I(t)$. This is also called the spin induced ellipticity of the transmitted light~\cite{book_Glazov}.

In the steady state, the average spin polarization is absent: $\braket{S_z(t)}=0$. Hereafter the angular brackets denote the quantum statistical average. In this case, the spin dynamics is characterized by the spin correlation function $\braket{S_z(t)S_z(t+\tau)}$~\cite{Zapasskii:13}. In the steady state it does not depend on $t$, so in the following we write it as $\braket{S_z(0)S_z(\tau)}$. From Eq.~\eqref{eq:Sz_I} one can see, that it is given by the correlation function of the circular polarization:
\begin{equation}
  \label{eq:I_corr}
  \braket{S_z(0)S_z(\tau)}\propto\braket{\Delta I(0)\Delta I(\tau)}.
\end{equation}
Detection of the correlation functions is known as the spin noise spectroscopy, and this is a particular case of the ellipticity measurement of the transmitted light.

The same time spin correlation functions for a single electron simply read~\cite{SinitsynReview}
\begin{equation}
  \label{eq:corr0}
  \braket{S_\alpha(0)S_\beta(0)}=\frac{\delta_{\alpha,\beta}}{4},
\end{equation}
where $\alpha,\beta=x,y,z$ and $\delta_{\alpha,\beta}$ is the Kronecker symbol. The correlation function $\braket{S_z(0)S_z(\tau)}$ is an even function of $\tau$ and for $\tau>0$ it satisfies the same equation of motion as $S_z(\tau)$~\cite{ll5_eng}. For this reason the spin correlation function gives the direct access to the spin dynamics and reveals the quantum Zeno effect even in the case, when the average spin polarization is absent.

In the next section we present the approaches to the calculation of the spin correlation function and establish the relation between them.

\section{Quantum Zeno effect}
\label{sec:results}
\subsection{Phenomenological approach}

Here we present a simple phenomenological approach to the quantum Zeno effect. Its validity for the system under study will be rigorously proven using the numerical and analytical approaches in the following subsections.

A single electron spin precession in the external magnetic field applied along the $x$ axis is generally described by the Hamiltonian
\begin{equation}
  {\cal H}_{0} = \hbar \Omega_{L}\sigma_{x}/2,
\end{equation}
here we introduce a vector composed of the spin Pauli matrices $\bm\sigma=(\sigma_x,\sigma_y,\sigma_z)$. Phenomenologically, the continuous weak measurement of the spin component $S_z$ is described by the following equation for the $2\times2$ spin density matrix $\rho(t)$~\cite{PhysRevA.36.5543,PRESILLA199695}:
\begin{equation}
  \label{eq:fen}
  \dot{\rho}(t) = -{\i\over \hbar}[{\cal H},\rho(t)]-{\lambda\over 2}[\sigma_{z},[\sigma_{z},\rho(t)]],
\end{equation} 
where $\lambda$ is the phenomenological ``measurement strength''. 
It can be associated with the quickly fluctuating magnetic field along the $z$ axis, which is provided by the randomly incoming and outcoming photons due to the dynamic Zeeman effect~\cite{Sussman2011,OpticalField}. This is quite general situation, so one can expect that this model correctly describes different ways of the spin measurement.

The electron spin is given by $\bm S(t)=\Tr[\rho(t)\bm\sigma/2]$, and from Eq.~\eqref{eq:fen} we obtain the kinetic equations
\begin{equation}
  \label{eq:dS}
  \dot{S}_x=-2\lambda S_x,
  \quad
  \dot{S}_y=-\Omega_L S_z-2\lambda S_y,
  \quad
  \dot{S}_z=\Omega_L S_y.
\end{equation}
These equations describe the electron spin precession around magnetic field and relaxation of the spin components $S_x$ and $S_y$, which do not commute with the observable $S_z$ due to the measurement back action.

From the two latter equations we obtain the two complex eigenfrequencies of the spin dynamics
\begin{equation}
  \label{eq:Omega_star}
  \Omega^*_\pm=\pm\sqrt{\Omega_L^2-\lambda^2}-\i\lambda.
\end{equation}
For the weak measurement strength $\lambda<\Omega_L$ the frequencies have opposite real parts and the same imaginary part $\lambda$. This describes the damped spin oscillations. With increase of the measurement strength, the absolute values of the real parts of the eigenfrequencies decrease and eventually become zero. For the strong measurement there are no spin oscillations and the spin decays like an overdamped oscillator.

From the solution of Eqs.~\eqref{eq:dS} we obtain
\begin{multline}
  S_{z}(t) = S_{z}(0)\left[\cos(\Omega t)+\frac{\lambda}{\Omega}\sin(\Omega t)\right]\e^{-\lambda t}\\
  +S_y(0)\frac{\Omega_{L}}{\Omega}\sin(\Omega t)e^{-\lambda t},
\end{multline}
where $\Omega = \sqrt{\Omega_{L}^{2}-\lambda^{2}}$ is the reduced spin precession frequency.

Using Eq.~\eqref{eq:corr0} and the fact that the spin correlation functions obey the analogous kinetic equations, we obtain the spin correlator~\cite{Liu2010,Bednorz}
\begin{equation}
  \label{eq:spin_corr}
	\expval{S_{z}(0)S_{z}(t)} ={1\over 4}\qty[\cos(\Omega t)+{\lambda\over \Omega}\sin(\Omega |t|)] e^{-\lambda |t|}.
\end{equation}
This expression shows that the quantum Zeno effect leads to the decrease of the spin precession frequency $\Omega$ and induces the spin relaxation. Qualitatively, this effect was observed for an ensemble of electrons in the planar microcavity~\cite{noise-trions}. Note, that it can also be called a watchdog effect~\cite{kraus1981measuring}.

The description of the quantum Zeno effect presented in this subsection is phenomenological. In the next subsections we prove that it can correctly describe the spin dynamics in the QD micropillar cavity and derive the microscopic expression for the phenomenological measurement strength $\lambda$.

\subsection{Numerical approach}

Here we present the numerical results obtained in the density matrix formalism based on the solution of the master equation~\eqref{eq:density_m} and compare them with the phenomenological theory described above.

The electron spin can not be measured directly, but only using the photons as described in Sec.~\ref{sec:model}. For this reason we study in this subsection the polarization of the transmitted light. The intensity of the transmitted $\sigma^\pm$ polarized light is proportional to the number of the corresponding photons in the cavity: $I_\pm(t)\propto c_\pm^\dag(t) c_\pm(t)$, so $\Delta I(t)\propto\Delta n(t)$, where
\begin{equation}
  \Delta n(t)=c_+^\dag(t)c_+(t)-c_-^\dag(t)c_-(t).
\end{equation}
Thus the correlator $\braket{S_z(0)S_z(t)}$ in Eq.~\eqref{eq:I_corr} is proportional to the correlator of $\Delta n(t)$ given by
\begin{multline}
  \label{eq:4c}
  \braket{\Delta n(0)\Delta n(t)}=2\bigg(\braket{c_+^\dag(0)c_+^\dag(t)c_+(t)c_+(0)}\\-\braket{c_+^\dag(0)c_-^\dag(t)c_-(t)c_+(0)}\bigg),
\end{multline}
where we used the fact, that $\Delta n(t)$ is zero on average and took into account the normal and time ordering of the operators~\cite{Carmichael}.

In Eq.~\eqref{eq:4c} the creation and annihilation operators in the Heisenberg representation are used. However, in the numerical calculations it is more convenient to switch to the the Schr\"odinger representation and to solve Eq.~\eqref{eq:density_m} for the time dependent density matrix.

The numerical calculation of Eq.~\eqref{eq:4c} consists of the three steps:

(i) We find the density matrix $\rho_0$ for the steady state.

(ii) Following the quantum regression theorem~\cite{Carmichael}, we calculate the density matrix of the system after a single $\sigma^+$ photon detection
\begin{equation}
  \label{eq:rho_0}
  \rho(0)=c_+\rho_0c_+^\dag.
\end{equation}
It represents the action of the two outer operators in Eq.~\eqref{eq:4c}.

(iii) We find the solution $\rho(t)$ of the master equation~\eqref{eq:density_m} with the initial condition~\eqref{eq:rho_0}. Then the correlators in Eq.~\eqref{eq:4c} are given by
\begin{equation}
  \braket{c_+^\dag(0)c_\pm^\dag(t)c_\pm(t)c_+(0)}=\Tr\left[c_+\rho(t)c_+^\dag\right].
\end{equation}
As a result, the correlator of the photon numbers reads
\begin{equation}
  \label{eq:n_corr}
  \braket{\Delta n(0)\Delta n(t)}=2\left\{\Tr\left[c_+^\dag c_+\rho(t)\right]-\Tr\left[c_-^\dag c_-\rho(t)\right]\right\}.
\end{equation}

This approach is exact and valid for the arbitrary relation between the parameters of the system. To speed up the numerical calculations, we consider the strong coupling regime and coherent excitation in the vicinity of the bare cavity resonance frequency:
\begin{equation}
  |\omega-\omega_0|,\varkappa,\gamma\ll g.
\end{equation}
In this limit we can neglect the polariton states and consider only the states, that are the product of the QD ground state and photon Fock states. In the numerical calculation we consider the states with up to $6$ photons and we checked that for $12$ photons the results are the same. This allows us to perform the exact calculations beyond the limit~\eqref{eq:weak_field}.

\begin{figure}
  \centering
  \includegraphics[width=\linewidth]{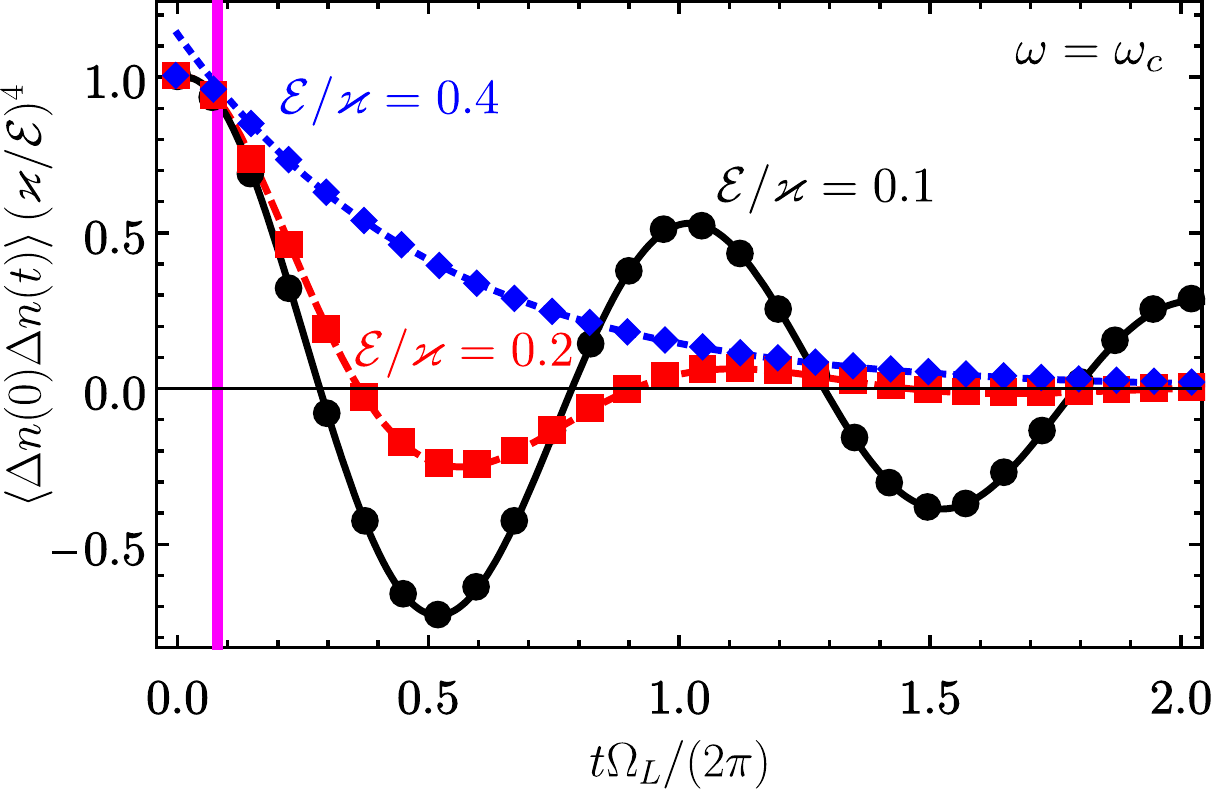}
  \caption{Dimensionless correlation function of the circular polarization degree of the transmitted light calculated numerically with the parameters $\omega=\omega_0$ and $\gamma=0$ in the limit $g\gg\varkappa$ for the different amplitudes of the incident light given in the labels with the corresponding color. The curves show the fits after Eq.~\eqref{eq:fit}. The vertical magenta line shows the time $t=5/\varkappa$.}
  \label{fig:Zeno_effect}
\end{figure}

In Fig.~\ref{fig:Zeno_effect} we show with the dots the correlator $\braket{\Delta n(0)\Delta n(t)}$ as a function of $t$ for the different amplitudes of the incident light $\mathcal E$. It indeed resembles the spin correlation function, Eq.~\eqref{eq:spin_corr}: it shows the damped oscillations. For this reason we fit $\braket{\Delta n(0)\Delta n(t)}$ with an expression
\begin{equation}
  \label{eq:fit}
  \Re\left(A\e^{-\i\Omega^*t}\right),
\end{equation}
where $A$ and $\Omega^*$ are the complex fit parameters. The fits are shown in Fig.~\ref{fig:Zeno_effect} by the curves of the corresponding color. One can see, that the fits nicely describe the numerical results for the time $t>5/\varkappa$. At shorter time scales the photon-photon interaction modifies the photon correlation functions~\cite{BackAction}.

\begin{figure}
  \centering
  \includegraphics[width=\linewidth]{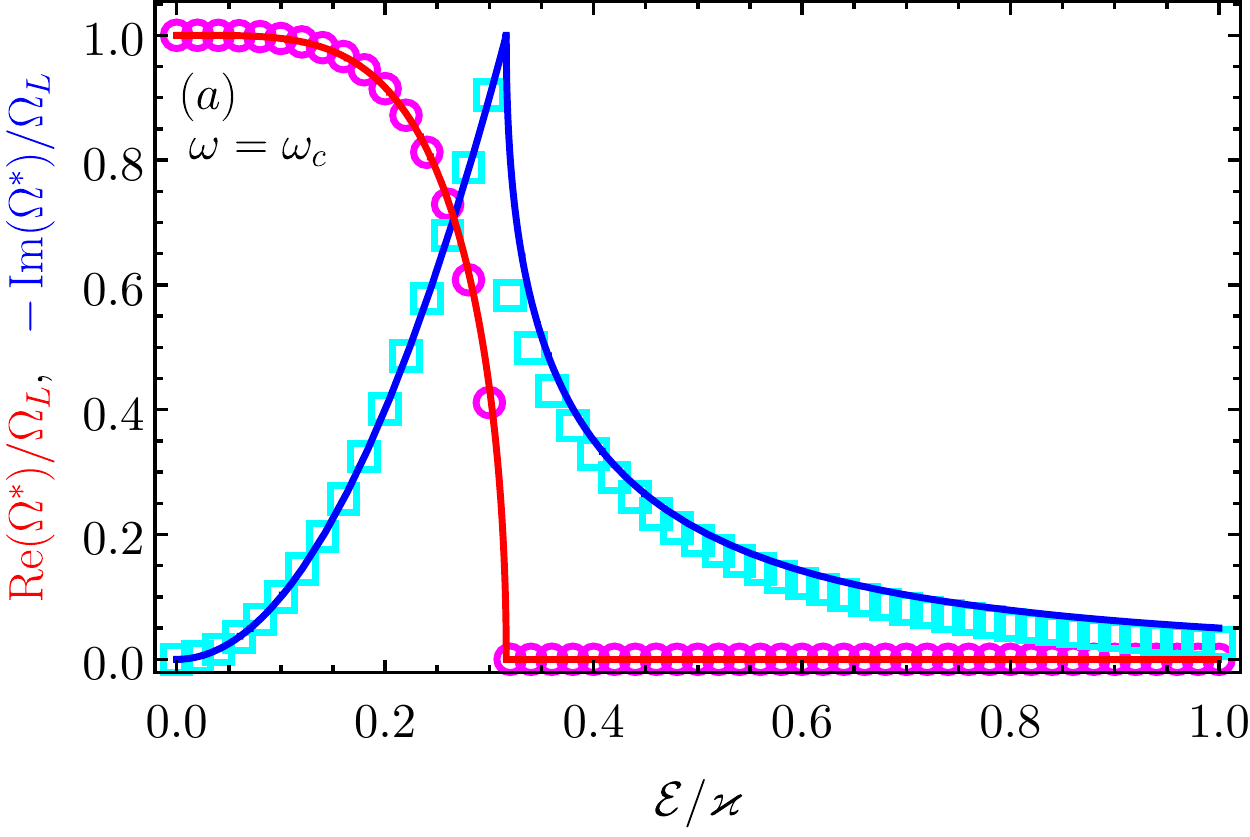}\\[2mm]
  \includegraphics[width=\linewidth]{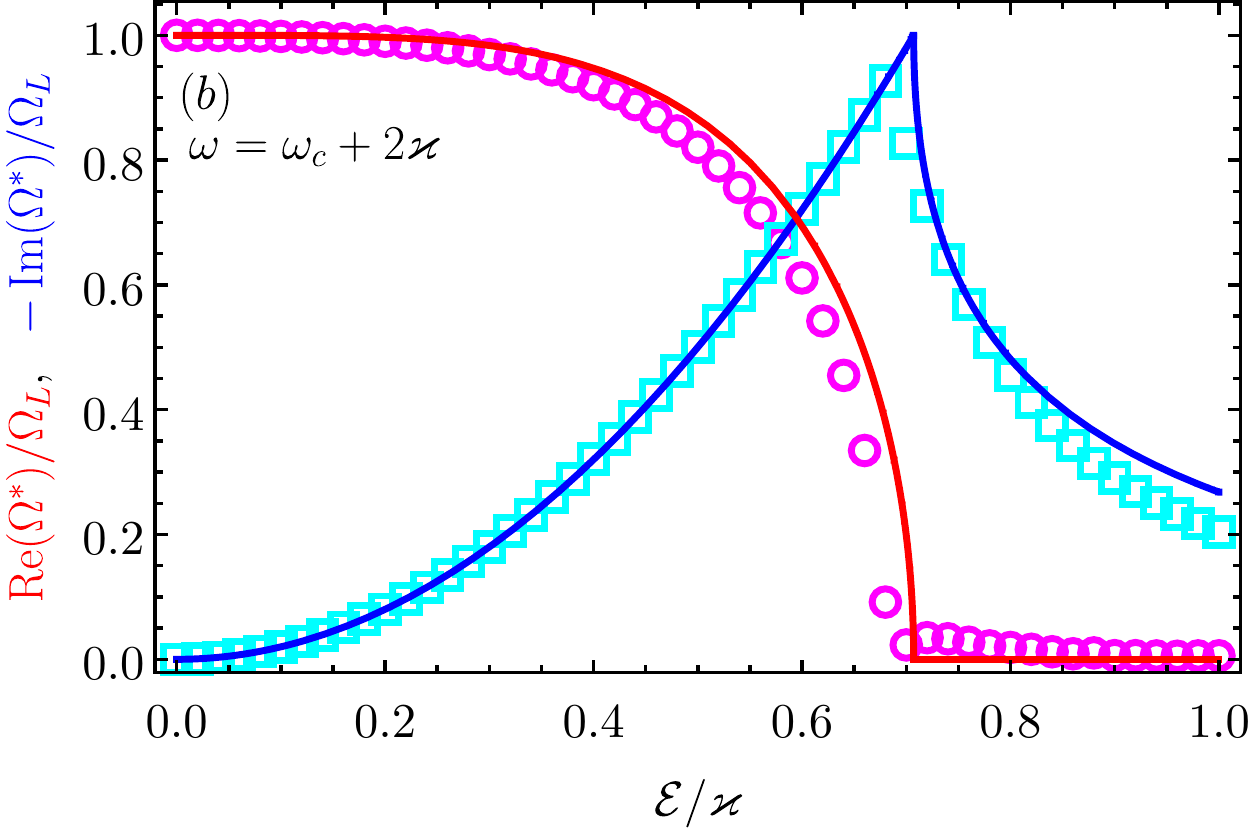}
  \caption{The spin precession frequency (red symbols) and spin relaxation rate (blue symbols) as functions of the amplitude of the incident light. The parameters of the calculation in (a) are the same as in Fig.~\ref{fig:Zeno_effect} and in (b) they are the same except for $\omega=\omega_c+2\varkappa$. The curves are calculated after Eqs.~\eqref{eq:Omega_star} and~\eqref{eq:quantum_lim} with the same parameters.}
  \label{fig:fit}
\end{figure}

We show the fit parameters $\Re(\Omega^*)$ and $-\Im(\Omega^*)$ in Fig.~\ref{fig:fit} by circles and squares, respectively, as functions of the amplitude of the incident light. One can see, that they show the similar behaviour for the different frequencies of the light [panels (a) and (b)]: The precession frequency $\Re(\Omega^*)$ monotonously decreases with increase of $\mathcal E$ and becomes zero after a certain threshold. The spin relaxation rate $-\Im(\Omega^*)$ first increases with increase of the intensity of the light, and then decreases after the threshold. In the limit of large power of the probe light, $\mathcal E\to\infty$, we obtain $\Omega^*=0$, which evidences the spin ``freezing'' due to the quantum Zeno effect under the strong continuous spin measurement.

We have checked, that the presented dependencies are qualitatively the same for any choice of the system parameters. We have also checked, that the relation $|\Omega^*|=\Omega_L$ is satisfied below the threshold in agreement with Eq.~\eqref{eq:Omega_star}. This shows that the phenomenological model of the quantum Zeno effect described in the previous subsection is valid for the system under study and that this effect can be observed in the intensity correlation functions of the transmitted light. In the next subsection we derive an analytical expression for the spin measurement strength $\lambda$.

\subsection{Analytical approach}
\label{sec:anal}

For the weak incident light and small magnetic field [Eq.~\eqref{eq:weak_field}], only a few lowest eigenstates are populated, so the master equation~\eqref{eq:density_m} can be solved analytically. This allows us to establish the applicability limits of the phenomenological approach and to calculate the measurement strength $\lambda$.

It is convenient to start the analysis from the simple limit, when there is no incident light, $\mathcal E=0$. In this case, only the two ground spin states are populated and only the four density matrix elements are nonzero. From Eq.~\eqref{eq:density_m} we find that
\begin{subequations}
  \label{eq:ground}
  \begin{equation}
    \label{eq:ground1}
    \dot{\rho}_{\uparrow, \uparrow}  = {\i\Omega_{L}\over 2}\qty(\rho_{\uparrow, \downarrow}-\rho_{\downarrow, \uparrow}),
  \end{equation}
  \begin{equation}
    \label{eq:ground2}
    \dot{\rho}_{\uparrow, \downarrow} = {\i\Omega_{L}\over 2}\qty(\rho_{\uparrow, \uparrow}-\rho_{\downarrow, \downarrow}).
  \end{equation}
\end{subequations}
Here the two electron spin states with $S_z=\pm1/2$ are labeled by the up and down arrows, respectively. The equation for ${\rho}_{\downarrow,\uparrow}$ can be obtained from Eq.~\eqref{eq:ground2} by the complex conjugation, and the equation for $\dot{\rho}_{\downarrow, \downarrow}$ follows from Eq.~\eqref{eq:ground1} and the normalization condition ${\rho}_{\uparrow, \uparrow}+{\rho}_{\downarrow, \downarrow}=1$. Both of them can be also obtained from Eqs.~\eqref{eq:ground} by the exchange of $\uparrow$ and $\downarrow$.

The electron spin components are given by
\begin{multline}
  \label{eq:spinden}
	S_{x} = \frac{\rho_{\uparrow,\downarrow}+\rho_{\downarrow,\uparrow}}{2},\qquad S_{y} = 
\i\frac{\rho_{\uparrow,\downarrow}-\rho_{\downarrow,\uparrow}}{2},\\ S_{z} = \frac{\rho_{\uparrow,\uparrow}-\rho_{\downarrow,\downarrow}}{2}.
\end{multline}
Using Eqs.~\eqref{eq:ground} we arrive at the phenomenological Eqs.~\eqref{eq:dS} with $\lambda=0$, as expected for the case, when the spin is not measured.

Eqs.~\eqref{eq:spinden} are valid for the case of the weak excitation also, when the two lowest states are mostly populated. To account for the finite excitation strength we use the perturbation theory and consider $\Omega_L/\varkappa$ and $\mathcal E/\varkappa$ as the small parameters. In the zeroth order, only the four density matrix elements discussed above are nonzero and their time derivatives vanish.

In the first order in $\mathcal{E}/\varkappa$ there are another twelve nonzero density matrix elements between the two ground states and the six excited states shown in Fig.~\ref{fig:eigen}. They can be found from the equations
\begin{subequations}
  \begin{multline}\label{eq:upD}
    \dot{\rho}_{\Downarrow_{u},\uparrow} =
    -\i(\omega_{0}+g)\rho_{\Downarrow_{u},\uparrow}-{\varkappa\over
      2}\qty(\rho_{\Downarrow_{u},\uparrow}-\rho_{\Downarrow_{l},\uparrow})-\\-{\gamma\over
      2}\qty(\rho_{\Downarrow_{u},\uparrow}+\rho_{\Downarrow_{l},\uparrow})-{\i{\cal
        E}e^{-\i\omega t}\over \sqrt{2}}\rho_{\downarrow,\uparrow},
  \end{multline}
  \begin{multline}
    \dot{\rho}_{\Downarrow_{l},\uparrow} =
    -\i(\omega_{0}-g)\rho_{\Downarrow_{l},\uparrow}-{\varkappa\over
      2}\qty(\rho_{\Downarrow_{l},\uparrow}-\rho_{\Downarrow_{u},\uparrow})-\\-{\gamma\over
      2}\qty(\rho_{\Downarrow_{u},\uparrow}+\rho_{\Downarrow_{l},\uparrow})+{\i{\cal
        E}e^{-\i\omega t}\over \sqrt{2}}\rho_{\downarrow,\uparrow},
  \end{multline}
  \begin{equation}\label{eq:updp}
    \dot{\rho}_{\downarrow+,\uparrow} = -\i\omega_{0}\rho_{\downarrow+,\uparrow}-\varkappa\rho_{\downarrow+,\uparrow}-{\i{\cal E}e^{-\i\omega t}\over \sqrt{2}}\rho_{\downarrow,\uparrow}.
  \end{equation}
\end{subequations}
Another nine equations can be obtained from these three by flipping the spins and by the Hermitian conjugation. 

The solution of these equations can be conveniently written using Eqs.~\eqref{eq:t01}:
\begin{subequations}
  \label{eq:first}
  \begin{equation}\label{eq:firsta}
    \rho_{\Downarrow_{l},\uparrow} = -\i \rho_{\downarrow,\uparrow}{{\cal E}e^{-\i\omega t}\over \sqrt{2}\varkappa}t_{1}\qty({g\over \omega-\omega_{0}+\i\gamma}-1),
  \end{equation}
  \begin{equation}\label{eq:firsts}
    \rho_{\Downarrow_{u},\uparrow} = -\i \rho_{\downarrow,\uparrow}{{\cal E}e^{-\i\omega t}\over \sqrt{2}\varkappa}t_{1}\qty({g\over \omega-\omega_{0}+\i\gamma}+1),
  \end{equation}
  \begin{equation}\label{eq:first0}
    \rho_{\downarrow+,\uparrow} = -\i{{\cal E}\over \varkappa}\rho_{\downarrow,\uparrow}t_{0}e^{-\i\omega t}.
  \end{equation}
\end{subequations}
One can check that the transmission coefficients $t_{0}$ and $t_1$ can be obtained from the density matrix in the first order in $\mathcal E$ from the following relation:
\begin{equation}
  \Tr\qty[\rho c_{\pm}] = -\i{{\cal E}e^{-\i\omega t}\over \varkappa}\qty(t_{1,0}\rho_{\uparrow,\uparrow}+t_{0,1}\rho_{\downarrow,\downarrow}),
  \label{eq:cpm_av}
\end{equation}
These relations show that the transmission coefficient of circularly polarize light is $t_0$ or $t_1$ when the helicity of the incident light is opposite or the same as the spin helicity.

The equations of motion for the electron spin can be obtained from the off-diagonal density matrix element between the ground states, see Eqs.~\eqref{eq:spinden}. In the two lowest orders of the perturbation theory we obtain
\begin{multline}\label{eq:updown}
	\dot{\rho}_{\uparrow, \downarrow} = {\i\Omega_{L}\over 2}\qty(\rho_{\uparrow, \uparrow}-\rho_{\downarrow, \downarrow})+\\+\sqrt{2}\varkappa\qty(\rho_{\Uparrow_{u},\downarrow+}+\rho_{\uparrow-,\Downarrow_{u}}-\rho_{\Uparrow_{l},\downarrow +}-\rho_{\uparrow-,\Downarrow_{l}})-\\-{\i{\cal E}\over \sqrt{2}}\left[e^{-\i\omega t}(\rho_{\uparrow,\Downarrow_{l}}-\rho_{\uparrow,\Downarrow_{u}}-\sqrt{2}\rho_{\uparrow,\downarrow+})\right.+\\+\left.e^{\i\omega t}(\rho_{\Uparrow_{u},\downarrow}-\rho_{\Uparrow_{l},\downarrow}+\sqrt{2}\rho_{\uparrow-,\downarrow})\right].
\end{multline}
Here the unknown density matrix elements can be found from another two equations in the second order in $\mathcal E/\varkappa$:
\begin{subequations}
  \begin{multline}\label{eq:sec1}
    \dot{\rho}_{\Uparrow_{u},\downarrow+} = -\i g
    {\rho}_{\Uparrow_{u},\downarrow+}+\i {\cal E}\qty(e^{\i\omega
      t}\rho_{\Uparrow_{u},\downarrow}-{e^{-\i\omega t}\over
      \sqrt{2}}\rho_{\uparrow,\downarrow+})-\\- {\gamma \over
      2}\qty(\rho_{\Uparrow_{u},\downarrow+}+\rho_{\Uparrow_{l},\downarrow+})-{\varkappa\over
      2}\qty(3\rho_{\Uparrow_{u},\downarrow+}-\rho_{\Uparrow_{l},\downarrow+}),
  \end{multline}
  \begin{multline}\label{eq:sec2}
    \dot{\rho}_{\Uparrow_{l},\downarrow+} = \i g
    {\rho}_{\Uparrow_{l},\downarrow+}+\i {\cal E}\qty(e^{\i\omega
      t}\rho_{\Uparrow_{l},\downarrow}+{e^{-\i\omega t}\over
      \sqrt{2}}\rho_{\uparrow,\downarrow+})-\\- {\gamma \over
      2}\qty(\rho_{\Uparrow_{l},\downarrow+}+\rho_{\Uparrow_{u},\downarrow+})-{\varkappa\over
      2}\qty(3\rho_{\Uparrow_{l},\downarrow+}-\rho_{\Uparrow_{u},\downarrow+}).
  \end{multline}
\end{subequations}
Their solution reads
\begin{subequations}
  \label{eq:2order_sol}
  \begin{equation}
    {\rho}_{\Uparrow_{u},\downarrow+} ={ \rho_{\uparrow,\downarrow}\over \sqrt{2}}\qty(\mathcal{E}\over\varkappa)^{2}\qty({g\over \omega-\omega_{0}+\i\gamma}+1)t_{0}^{*}t_{1},
  \end{equation}
   \begin{equation}
    {\rho}_{\Uparrow_{l},\downarrow+} ={ \rho_{\uparrow,\downarrow}\over \sqrt{2}}\qty(\mathcal{E}\over\varkappa)^{2}\qty({g\over \omega-\omega_{0}+\i\gamma}-1)t_{0}^{*}t_{1}.
  \end{equation}
\end{subequations}
Another two density matrix elements in Eq.~\eqref{eq:updown}, $\rho_{\uparrow-,\Downarrow_{u}}$ and $\rho_{\uparrow-,\Downarrow_{l}}$ are obtained from these two by the flip of helicities and Hermitian conjugation.

Finally, substituting Eqs.~\eqref{eq:2order_sol} and~\eqref{eq:first} in Eq.~\eqref{eq:updown} we obtain
\begin{equation}
  \dot{\rho}_{\uparrow, \downarrow} = {\i\Omega_{L}\over 2}\qty(\rho_{\uparrow, \uparrow}-\rho_{\downarrow, \downarrow})-2\lambda \rho_{\uparrow,\downarrow},
\end{equation}
in agreement with the phenomenological Eqs.~\eqref{eq:dS} for $S_x$ and $S_y$. Here the measurement strength is given by~\footnote{It can be also compactly written as $\lambda=(\mathcal E^2/\varkappa)\Re\qty(t_{0}+t_{1}-2t_{0}t_{1}^{*})$.}
\begin{equation}
\label{eq:Zeno_rate}
  \lambda=\frac{\mathcal E^2}{\varkappa}|t_0-t_1|^2\left(1+\frac{\gamma\left[(\omega-\omega_0)^2+\varkappa^2\right]}{g^2\varkappa}\right).
\end{equation}
This is the main result of this section.

In a similar way one can consider the equation of motion for $\rho_{\uparrow, \uparrow}$ in the second order of the perturbation theory. In this case the terms with $\mathcal E$ and $\varkappa$ cancel each other, so one simply obtains
\begin{equation}
  \dot{\rho}_{\uparrow, \uparrow}  = {\i\Omega_{L}\over 2}\qty(\rho_{\uparrow, \downarrow}-\rho_{\downarrow, \uparrow}),
\end{equation}
which agrees with Eq.~\eqref{eq:dS} for $S_z$.

In the absence of trion nonradiative decay, $\gamma=0$, the measurement strength, Eq.~\eqref{eq:Zeno_rate}, is simply given by:
\begin{equation}
  \label{eq:quantum_lim}
  \lambda=\frac{\mathcal E^2}{\varkappa}|t_0-t_1|^2.
\end{equation}
In Fig.~\ref{fig:fit} we use this relation between $\lambda$ and $\mathcal E$ along with Eq.~\eqref{eq:Omega_star} to calculate the complex spin precession frequency. One can see that it agrees with the fit of the numerical results. The difference between them is mainly due to the moderate ratio $\Omega_L/\varkappa=0.1$, which is assumed to be very small in the derivation of Eq.~\eqref{eq:quantum_lim}.

Beyond the limit~\eqref{eq:weak_field} the excited states should be considerably populated in order to suppress the electron spin precession in the magnetic field. In this case the phenomenological theory breaks down, so the quantum Zeno effect can be described only microscopically taking into account the parameters of the trion spin dynamics such as the heavy hole $g$ factor and spin relaxation time~\cite{noise-trions}.

Noteworthy, we find that the dependence of $\lambda$ on $\omega-\omega_0$ and $g$ in Eq.~\eqref{eq:quantum_lim} reduces to the amplitude transmission coefficients only. This suggests, that the measurement strength in this limit can be obtained without detailed analysis of the excited states. This is done in the next section.

\section{Quantum nondemolition spin measurement}
\label{sec:informatics}

In this section we discuss the quantum Zeno effect in the QD micropillar cavity from the perspective of the quantum informatics. In the previous section we demonstrated, that the quantum Zeno effect for weak fields is described by Eqs.~\eqref{eq:dS} with the measurement strength given by Eq.~\eqref{eq:Zeno_rate}.

The relaxation of the spin components $S_x$ and $S_y$ arises due to the spin measurement back action. Generally, the relaxation (dephasing) rate of the off-diagonal density matrix element, $\Gamma_{\rm deph}$, is greater than or equal to the measurement rate, $\Gamma_{\rm meas}$~\cite{devoret2000amplifying,PhysRevB.64.165310,Clerk}:
\begin{equation}
  \label{eq:noneq}
  \Gamma_{\rm deph}\ge\Gamma_{\rm meas}.
\end{equation}
They equal to each other in the quantum limit. Note the difference between the measurement rate, $\Gamma_{\rm meas}$, which we rigorously define below and the measurement strength $\lambda$, introduced in the previous section.

The inequality~\eqref{eq:noneq} holds for any external magnetic field, and in particular for $\Omega_L=0$. In this case the relaxation of the spin component $S_z$ is absent, and the quantum nondemolition measurement is realized. In this section we consider this limit and identify the conditions when the quantum limit for the spin measurement is reached.

\subsection{Optical spin measurement rate}

To recall the definition of the measurement rate $\Gamma_{\rm meas}$ we assume, that the spin is initially prepared in one of the states $S_z=\pm1/2$, e.g. by a short circularly polarized pump pulse~\cite{singleSpin}. Consequently, the spin orientation can be determined using the continuous probe light, as discussed in Sec.~\ref{sec:model}. In the absence of magnetic field the measurement is quantum nondemolition, so $S_z$ does not change. After a short measurement time $t$, after a few photons are detected, one can not say for sure what is the electron spin state, but one can calculate the two conditional probabilities $P_\pm(t)$ for the spin-up and spin-down states, respectively.

The quantum statistical average
\begin{equation}
  \mathcal I(t)=-\sum_\pm\overline{P_\pm(t)\ln[P_\pm(t)]},
  \label{eq:entropy}
\end{equation}
yields the ``entropy'', which characterizes our knowledge about the spin direction. Note, that $\mathcal I(t)$ does not describe the real entropy of the system, which is zero, because the electron is actually completely spin polarized. The measurement rate is defined by
\begin{equation}
  \label{eq:meas_def}
  \Gamma_{\rm meas}=-\dot{\mathcal I}(0),
\end{equation}
and represents the ``entropy'' decrease rate at the beginning of the spin measurement.

In the previous section we considered the spin measurement by the detection of the intensities of $\sigma^+$ and $\sigma^-$ polarized components of the transmitted light. Their dependence on the spin direction, Eq.~\eqref{eq:T_pm}, allows one to deduce the spin direction from the circular polarization of the transmitted light. Generally, the polarization of the light is described by the vectors of the Stokes parameters $\bm\xi_\pm$ for the cases of spin-up and spin-down electron, respectively~\cite{ll2_eng,ivchenko05a}.

\begin{figure}
  \centering
  \includegraphics[width=\linewidth]{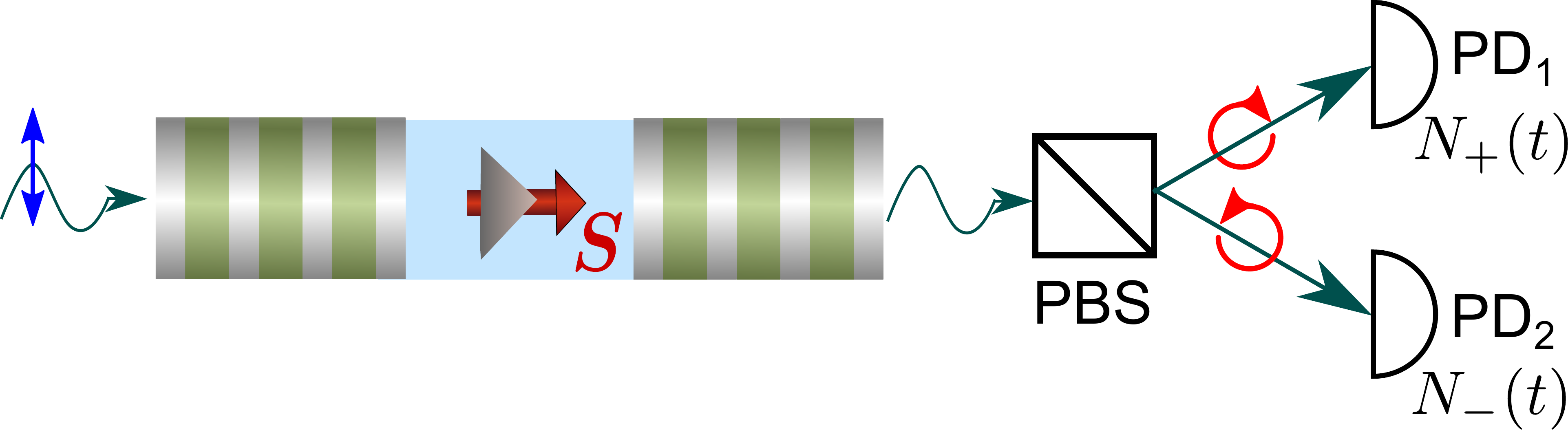}
  \caption{Scheme of the quantum nondemolition spin measurement by linearly polarized light. The magnetic field is absent. The transmitted light is separated into two orthogonally polarized components by the polarizing beam splitter (PBS), and then the photons in the two polarizations are independently counted using the photodetectors (PDs).}
  \label{fig:measurement}
\end{figure}

The amplitude transmission coefficients of $\sigma^\pm$ light are given similarly to Eq.~\eqref{eq:T_pm} by
\begin{equation}
  \label{eq:tpm}
  t_\pm=\frac{t_0+t_1}{2}\pm(t_1-t_0)S_z,
\end{equation}
see also Eqs.~\eqref{eq:cpm_av}. In the canonical basis~\cite{Varshalovich} the circular components of the electric field are given by $E_\pm=(\mp E_x+\i E_y)/\sqrt{2}$, so the Stokes parameters can be expressed as
\begin{multline}
  \label{eq:Stokes_true}
  \xi_{1,\pm}=\pm\frac{2\Im(t_1t_0^*)}{|t_1^2|+|t_0^2|},
  \qquad
  \xi_{2,\pm}=\pm\frac{|t_1^2|-|t_0^2|}{|t_1^2|+|t_0^2|},\\
  \xi_{3,\pm}=-\frac{2\Re(t_1t_0^*)}{|t_1^2|+|t_0^2|}.
\end{multline}
Thus, the spin polarization can be measured using both the Faraday rotation and the ellipticity of the transmitted light, which are proportional to $\xi_1$ and $\xi_2$, respectively. To determine one of them, the light transmitted through the cavity can be analyzed using the polarizing beam splitted and two photodetectors, as shown in Fig.~\ref{fig:measurement}. Note, that the total intensity of the transmitted light being proportional to $T_++T_-$ is independent of $S_z$, see Eq.~\eqref{eq:T_pm}.

To maximize the detection sensitivity one can adjust the polarizations separated by the beam splitter to measure a combination of the Faraday rotation and ellipticity signals. Generally, by the appropriate rotation of the polarization basis of the Poincare sphere one can obtain the Stokes parameters of the form
\begin{equation}
  \label{eq:Stokes_simple}
  \xi_{1',\pm}=0,
  \qquad
  \xi_{2',\pm}=\pm\xi,
  \qquad
  \xi_{3',+}=\xi_{3',-}.
\end{equation}
Here the primes denote the rotated axes. In this basis the spin state is distinguished by the second Stokes parameter only, which is the circular polarization degree. From Eqs.~\eqref{eq:Stokes_true} one can see, that
\begin{equation}
  \label{eq:xi}
  \xi=\sqrt{\xi_{1,\pm}^2+\xi_{2,\pm}^2}=\frac{|t_1^2-t_0^2|}{|t_1^2|+|t_0^2|}.
\end{equation}
As expected, the Stokes parameters are different only if the transmission coefficients $t_0$ are $t_1$ are different.

In what follows we assume that $\xi_{2'}$ is measured as shown in Fig.~\ref{fig:measurement}. We denote the numbers of the detected circularly polarized photons after the measurement time $t$ as $N_\pm(t)$, see Fig.~\ref{fig:measurement}. If they are large, the Stokes parameter is given by
\begin{equation}
  \label{eq:xi2prime}
  \xi_{2'}=\frac{N_+-N_-}{N_++N_-}=\pm\xi.
\end{equation}
However, if $N_\pm$ are small, this equation yields only an estimate for $\xi_{2'}$, which is a random number and can differ from $\pm\xi$.

\subsection{Entanglement between electron and photon}

The spin measurement through the photon detection can be viewed as a result of the entanglement between electron and photon spins~\cite{DeGreve2012,Gao:2012fk,PhysRevLett.110.167401}. The electron and photon represent in this case the two qubits. The photon detection destroys the entanglement, which leads to the dephasing of the electron spin. The frequent photon detection leads to the quick relaxation of the off-diagonal spin density matrix elements, which causes the quantum Zeno effect. So the measurement strength is determined by the degree of the spin-photon entanglement.

To put it on a quantitative basis, we write down the two-particle wave function of an electron and a single transmitted photon:
\begin{equation}
  \label{eq:psi_e-ph}
  \ket{\Psi_{\rm{e-ph}}}=\frac{\left(\chi_+t_1,\chi_+t_0,\chi_-t_0,\chi_-t_1\right)}{\sqrt{|t_1^2|+|t_0^2|}}.
\end{equation}
Here we use the basis states $\ket{\uparrow,\sigma^+}$,  $\ket{\uparrow,\sigma^-}$,  $\ket{\downarrow,\sigma^+}$,  $\ket{\downarrow,\sigma^+}$, with an arrow denoting the electron spin state and $\sigma^\pm$ denoting the photon polarization. The components of the electron spinor are denoted as $\chi_\pm$. Eq.~\eqref{eq:psi_e-ph} represents a pure state, so its concurrence is given by~\cite{PhysRevLett.80.2245}
\begin{equation}
  \mathcal C=\left|\left<\Psi_{\rm{e-ph}}\middle|\mathcal T\middle|\Psi_{\rm{e-ph}}\right>\right|,
\end{equation}
where $\mathcal T$ denotes the time inversion (multiplication by the second Pauli matrices of both qubits and complex conjugation). The straightforward calculation yields
\begin{equation}
  \label{eq:C}
  \mathcal C=2\xi|S_x+\i S_y|. 
\end{equation}
This establishes the relation between the concurrence and the difference of the Stokes parameters $\xi_{2',\pm}$. The spin components $S_x$ and $S_y$ in Eq.~\eqref{eq:C} reflect the fact, that the destruction of the entanglement leads to the relaxation of the off-diagonal density matrix elements, see Eqs.~\eqref{eq:fen} and~\eqref{eq:dS}.

\subsection{Calculation of the measurement rate}

To calculate the measurement rate we consider a short measurement time $t$, when the average number of the detected photons $\overline{N(t)}=\overline{N_+(t)}+\overline{N_-(t)}$ is small: $\overline{N(t)}\ll 1$. In this case, we can neglect the probability to detect two photons and consider zero or one detected photon only.

In the case of $N_+=1$ and $N_-=0$, the conditional probabilities of spin-up and spin-down states are given by
\begin{equation}
  \label{eq:Wpm}
  P_\pm=\frac{1\pm\xi}{2},
\end{equation}
see Eq.~\eqref{eq:xi2prime}. In the case of $N_+=0$, $N_-=1$, the $\pm$ sign in the right hand side should be flipped. And in the trivial case of $N_+=N_-=0$ one has $P_\pm=1/2$, since there is no ground for the spin direction estimation.

As an example, if $\xi=1$ then the transmitted light is always circularly polarized and its polarization is $\sigma^\pm$ (in the rotated basis) for $S_z=\pm1/2$, respectively. In this case, a single photon detection strictly yields the electron spin orientation: either $P_+$ or $P_-$ equals to unity as follows from Eq.~\eqref{eq:Wpm}.

In the opposite limit of $\xi\ll 1$, the polarizations of the transmitted light for $S_z=\pm1/2$ are very similar, and after a single photon detection $P_\pm\approx1/2$ in Eq.~\eqref{eq:Wpm}, so it is difficult to say what is the actual electron spin state.

Generally, Eq.~\eqref{eq:Wpm} along with the definition~\eqref{eq:entropy} yields
\begin{equation}
  \mathcal I(t)=\ln(2)-\overline{N(t)}\left[\frac{1+\xi}{2}\ln(1+\xi)+\frac{1-\xi}{2}\ln(1-\xi)\right].
\end{equation}
This leads to the measurement rate [Eq.~\eqref{eq:meas_def}]
\begin{equation}
  \label{eq:Gamma_t}
  \Gamma_{\rm meas}=\left[\frac{1+\xi}{2}\ln(1+\xi)+\frac{1-\xi}{2}\ln(1-\xi)\right]\overline{\dot{N}},
\end{equation}
where $\overline{\dot{N}}$ is the average flux of the transmitted photons. This expression is general. It is valid for arbitrary $\xi$ and can be used to calculate the measurement rate for any optical spin measurement.

The photon flux is given by
\begin{equation}
  \overline{\dot{N}}=2\varkappa_2\braket{c_+^\dag c_++c_-^\dag c_-},
\end{equation}
where we recall that $\varkappa_2$ is the photon amplitude decay rate through the right mirror of the cavity (the light is assumed to be incident at the left one). Making use of the steady state density matrix found in Sec.~\ref{sec:anal}, we obtain
\begin{equation}
  \label{eq:N_cav}
  \overline{\dot{N}}=2\varkappa_2\frac{\mathcal E^2}{\varkappa^2}\left(|t_0^2|+|t_1^2|\right).
\end{equation}
Now the substitution of this expression expression along with Eq.~\eqref{eq:xi} in Eq.~\eqref{eq:Gamma_t} yields the spin measurement rate $\Gamma_{\rm meas}$.

It should be compared with the dephasing rate, $\Gamma_{\rm deph}$, the rate of the decay of the off-diagonal spin density matrix element:
\begin{equation}
  \label{eq:deph_lambda}
  \Gamma_{\rm deph}=2\lambda,
\end{equation}
see Eqs.~\eqref{eq:dS} and~\eqref{eq:spinden}. Here $\lambda$ is given by Eq.~\eqref{eq:Zeno_rate}.

\begin{figure}
  \centering
  \includegraphics[width=\linewidth]{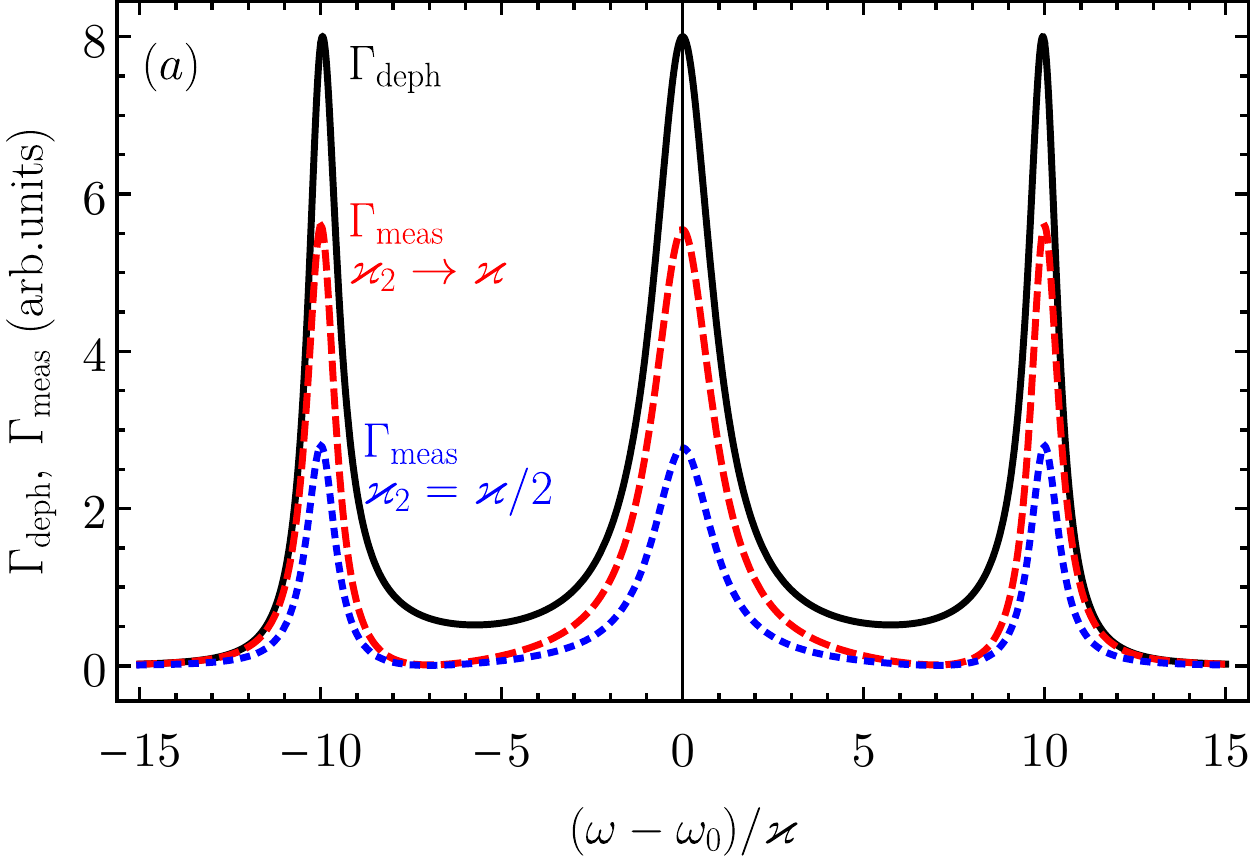}\\[2mm]
  \includegraphics[width=\linewidth]{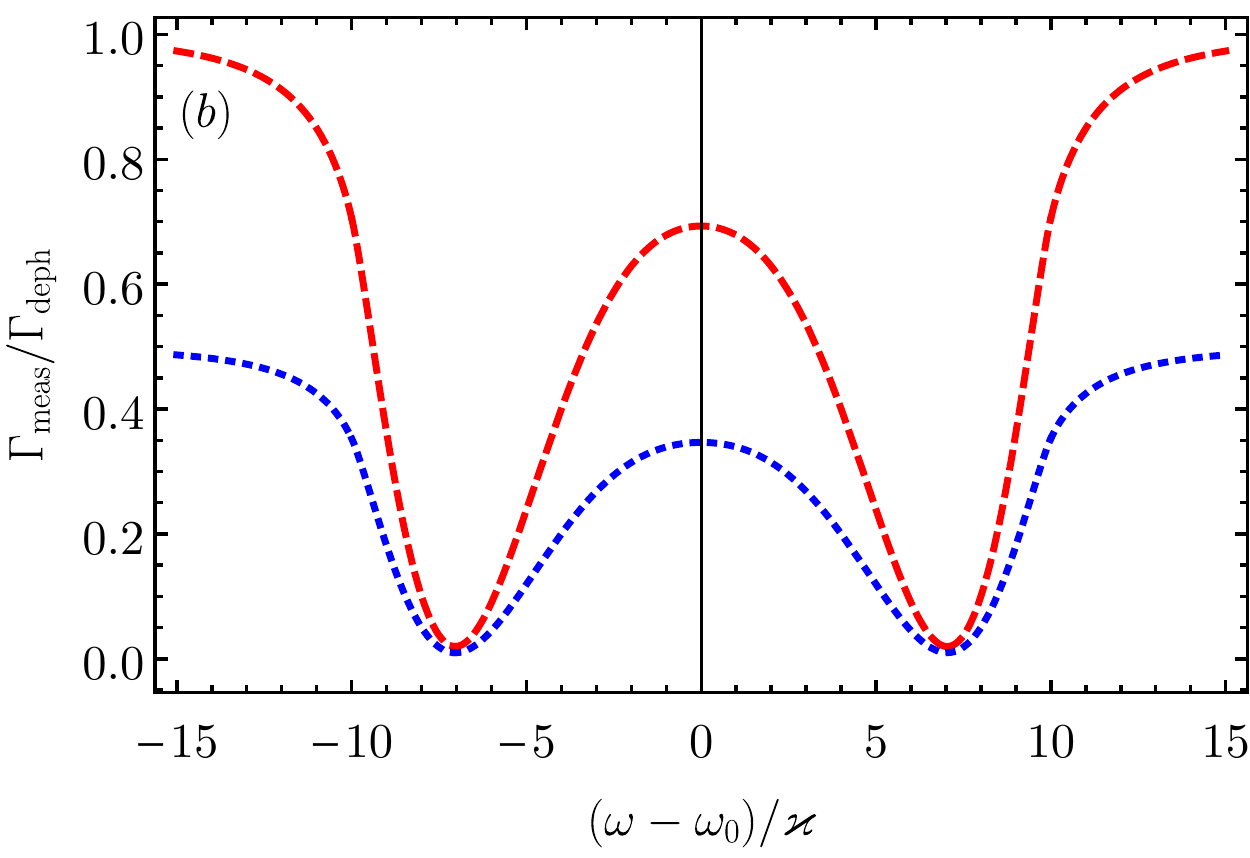}
  \caption{(a) Spin measurement and dephasing rates, calculated after Eqs.~\eqref{eq:Gamma_t} (red dashed and blue dotted curves) and~\eqref{eq:deph_lambda} (black solid curve), respectively, for $\gamma=0$ and $g/\varkappa=10$. For the blue dotted and red dashed curves $\varkappa_2=\varkappa/2$ and $\varkappa_2\to\varkappa$, respectively. (b) Ratio of the measurement and dephasing rates for the same parameters shown with the same colors.}
  \label{fig:rates}
\end{figure}

In Fig.~\ref{fig:rates}(a) we compare the spin measurement and dephasing rates as functions of the probe frequency. The both rates show the three peaks at the same frequencies as the transmission coefficient: bare cavity and polariton frequencies. One can see, that the dephasing rate is always larger, than the measurement rate, so the inequality~\eqref{eq:noneq} is satisfied.

The trion decay with the rate $\gamma$ increases the dephasing rate in Eq.~\eqref{eq:Zeno_rate}, so in Fig.~\ref{fig:rates} we consider the limit $\gamma=0$. Moreover, it follows from Eq.~\eqref{eq:N_cav} that the detected photon flux is proportional to the ratio $\varkappa_2/\varkappa$, so the more photons escape the cavity through the right mirror the higher the measurement rate. The blue dotted and red dashed curves demonstrate that the measurement rate is two times larger in the limit $\varkappa_2\to\varkappa$ than in the case of symmetric cavity, $\varkappa_2=\varkappa/2$. However, even if all the photons escape the cavity through the right mirror and are captured by the photodetectors, the measurement rate is still considerably smaller, than the dephasing rate, as shown in Fig.~\ref{fig:rates}(b). Only for the strongly detuned probe light, the measurement rate approaches the dephasing rate.

\subsection{Reaching quantum limit}

To summarize the previous subsection, the optimization of the structure parameters, $\gamma=0$ and $\varkappa_2\to\varkappa$, does not allow one to reach the quantum limit. This can be explained using the expression for the difference of the Stokes parameters~\eqref{eq:xi}. For example, if $t_0=1$ and $t_1=-1$ then the transmitted light is always polarized linearly perpendicular to the polarization of the incident light. The Stokes parameters are the same for spin up and spin down electron, so the detection scheme shown in Fig.~\ref{fig:measurement} does not allow one to determine the electron spin state. However, the phase of the transmitted light is opposite for the two spin states. This phase can be measured using the homodyne detection. The homodyne detection is often used to reach the quantum limit~\cite{Clerk}, but up to date it was not clear how it changes the measurement rate.

\begin{figure}[t]
  \centering
  \includegraphics[width=1\linewidth]{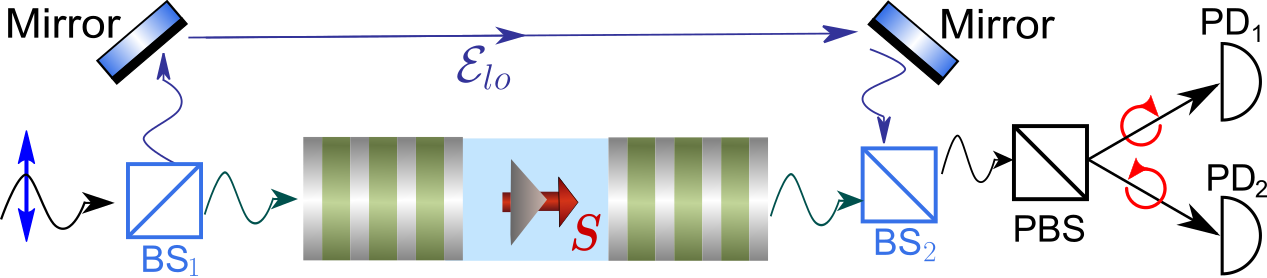}
  \caption{Scheme of the homodyne nondemolition spin measurement. Linearly polarized light is separated into two beams by the beam splitter BS$_{1}$. After one of them passes through the cavity, they are mixed by the beam splitter BS$_2$. The photon counting scheme is the same as in Fig.~\ref{fig:measurement}.}
  \label{fig:homo}
\end{figure}

In the homodyne detection scheme, Fig.~\ref{fig:homo}, the light emitted from the cavity interferes with the field of local oscillator, which has the fixed phase relative to the light incident at the cavity. The total amplitudes of the circularly polarized components of light before the polarizing beam splitted have the form
\begin{equation}
  E_\pm\propto t_\pm\mathcal E+\mathcal E_{lo},
\end{equation}
where $\mathcal E_{lo}$ is proportional to the amplitude of the local oscillator field (the effective decrease of the transmission coefficients $t_\pm$ due to the beam splitter BS$_2$ in Fig.~\ref{fig:homo} can be accounted for by the renormalization of $\mathcal E_{lo}$). In this case, the above theory is valid provided the effective transmission coefficients
\begin{equation}
  \tilde t_{\pm}=t_\pm+t_{lo}
\end{equation}
are used instead of $t_\pm$ with $t_{lo}=\mathcal E_{lo}/\mathcal E$. For the large amplitude of the local oscillator, $t_{lo}\gg 1$, the difference between the Stokes parameters is small:
\begin{equation}
  \xi=\frac{|t_1-t_0|}{|t_{lo}|}\ll 1,
\end{equation}
so the measurement rate Eq.~\eqref{eq:Gamma_t} can be written as
\begin{equation}
  \label{eq:small_xi}
  \Gamma_{\rm meas}=\frac{\xi^2}{2}\overline{\dot{N}}.
\end{equation}
However, the photon flux is $2|t_{lo}^2|/(|t_1^2|+|t_0^2|)$ times larger than without the homodyne detection [Eq.~\eqref{eq:N_cav}]. So the total measurement rate for the homodyne detection is
\begin{equation}
  \Gamma_{\rm meas}^{\rm homo}=2|t_1-t_0|^2\varkappa_2\frac{\mathcal E^2}{\varkappa^2}.
\end{equation}
From the comparison with Eqs.~\eqref{eq:deph_lambda} and~\eqref{eq:Zeno_rate} one can see that for $\gamma=0$ and $\varkappa_2=\varkappa$ the quantum limit is reached at all frequencies of the probe light.

\section{Spin statistics}
\label{sec:statistics}

The gain of the information about the electron spin unavoidably leads to the quantum Zeno effect and modifies the whole electron spin statistics. In this section we return to the case, when external transverse magnetic field is applied to the system.
Vanishing of the spin precession frequency with increase of the measurement strength shown in Fig.~\ref{fig:fit} evidences the qualitative change of the regime of the spin dynamics.

In particular, the modification of the spin statistics can be seen in the spin correlation functions beyond the second order~\cite{Sinitsyn-Correlators}. In the system under study the third order correlator vanishes, because the spin is zero on average. So in this section we analyze the fourth order spin correlation function.

For the classical Gaussian noise the high order correlators reduce to the sum of products of the second order correlation functions similarly to the Wick's theorem. Generally, to describe the nontrivial contribution to the fourth order correlation function, the cumulant is introduced according to
\begin{multline}
  \label{eq:C4_def}
  C_4(\tau_1,\tau_2,\tau_3)=\left<\left\{S_z(0)\left\{S_z(\tau_1)\left\{S_z(\tau_2)S_z(\tau_3)\right\}_s\right\}_s\right\}_s\right>
  \\
  -\left<\left\{S_z(0)S_z(\tau_1)\right\}_s\right>\left<\left\{S_z(\tau_2)S_z(\tau_3)\right\}_s\right>
  \\
  -\left<\left\{S_z(0)S_z(\tau_2)\right\}_s\right>\left<\left\{S_z(\tau_1)S_z(\tau_3)\right\}_s\right>
  \\
  -\left<\left\{S_z(0)S_z(\tau_3)\right\}_s\right>\left<\left\{S_z(\tau_1)S_z(\tau_2)\right\}_s\right>.
\end{multline}
Here $\left\{AB\right\}_s=(AB+BA)/2$ and it is assumed that $0<\tau_1<\tau_2<\tau_3$. 

For a single spin, the nonzero fourth order cumulant stems from the ``quantum nature'' of spin: it takes only one of the two eigenvalues $S_z=\pm1/2$ at the four moments $0$, $\tau_1$, $\tau_2$, and $\tau_3$. Usually, the fourth order spin correlation function is hardly accessible experimentally~\cite{PhysRevB.101.235416}, but in the system under study it can be easily determined from the photon counting statistics, as described in Sec.~\ref{sec:model} for the second order correlation function.

The higher order spin correlators can be calculated, for example, using the path integrals~\cite{Sinitsyn-Correlators,Bednorz} or Ito calculus~\cite{doi:10.1080/00107510601101934,PhysRevB.98.205143}. In our case of Markovian spin dynamics, one can simply consider all $16$ possible spin states at the four time moments and calculate the probabilities of the corresponding quantum spin trajectories using the Lindblad equation~\eqref{eq:fen}.

One can show that generally for a single spin, the first two terms in Eq.~\eqref{eq:C4_def} cancel each other~\footnote{The details of this derivation will be published elsewhere.}, so
\begin{multline}
  \label{eq:C42}
  C_4(\tau_1,\tau_2,\tau_3)=
  -\left<S_z(0)S_z(\tau_2)\right>\left<S_z(\tau_1)S_z(\tau_3)\right>
  \\
  -\left<S_z(0)S_z(\tau_3)\right>\left<S_z(\tau_1)S_z(\tau_2)\right>.
\end{multline}
This allows us to calculate the fourth order spin correlation function using Eq.~\eqref{eq:spin_corr}.

The fourth order correlator depends on the three time intervals, and its Fourier transform depends on the three frequencies. For simplicity, it is often reduced to the bispectrum which is given by
\begin{equation}
  B(\omega_1,\omega_2)=\iint\d\tau_1\d\tau_2\d\tau\e^{\i\omega_1\tau_1+\i\omega_2\tau_2}C_4(\tau_1,\tau,\tau+\tau_2).
\end{equation}
Qualitatively it reflects the degree of the correlation between the spin noise at frequencies $\omega_1$ and $\omega_2$.

\begin{figure}
  \centering
  \includegraphics[width=\linewidth]{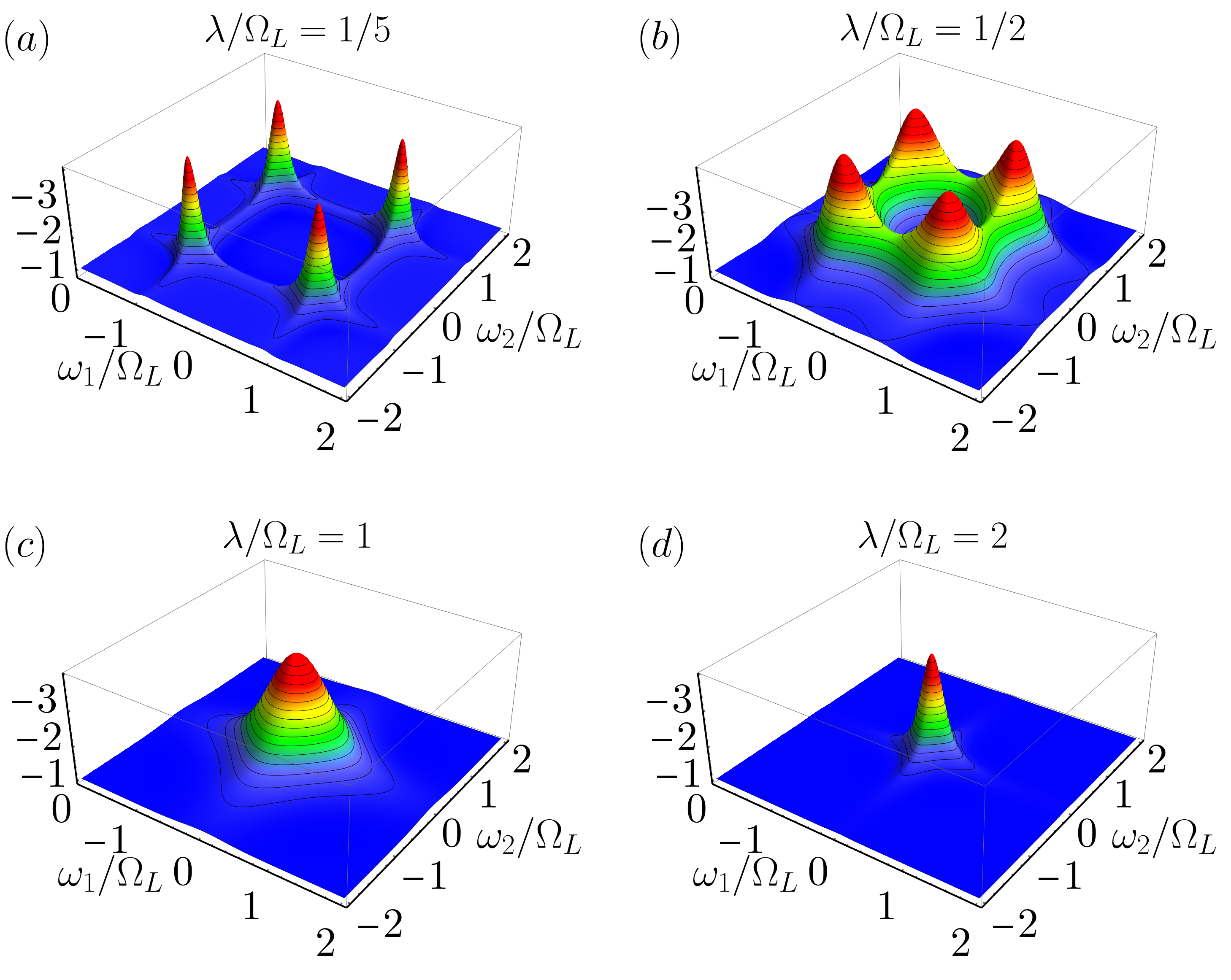}
  \caption{Spin noise bispectrum $B(\omega_1,\omega_2)$ in arbitrary units calculated after Eq.~\eqref{eq:C42} for the different measurement strengths given in the labels.}
  \label{fig:bispectrum}
\end{figure}

The modification of the bispectrum with increase of the measurement strength is shown in Fig.~\ref{fig:bispectrum}. Note that the vertical axis goes from top to bottom, so the bispectrum is mostly negative. It is always an even function of $\omega_1$ and $\omega_2$.

Under the weak measurement, $\lambda\ll\Omega_L$ [panel (a)], the bispectrum consists of four peaks at the frequencies $\omega_{1,2}=\pm\Omega_L$. For the positive frequencies the bispectrum has the form
\begin{equation}
  B(\omega_1,\omega_2)=\frac{\lambda[(\delta_1+\delta_2)^2+4\lambda^2](\delta_1\delta_2-\lambda^2)}{16(\lambda^2+\delta_1^2)^2(\lambda^2+\delta_2^2)^2},
\end{equation}
where $\delta_{1,2}=\omega_{1,2}-\Omega_L$.

With increase of the measurement strength the peaks become broader and shift to the lower frequencies along $|\omega_1|=|\omega_2|$ directions. The peaks start to overlap as shown in panel (b) and eventually merge at $\omega_1=\omega_2=0$ at the phase transitions, $\lambda=\Omega_L$ [panel (c)]. Note, that there is no abrupt change in the bispectrum as in the case of the second order phase transition described by the continuously changing order parameter.

Under the strong spin measurement $\lambda\gg\Omega_L$ the spin noise bispectrum has the form
\begin{equation}
  B(\omega_1,\omega_2)=\frac{\gamma}{2}\frac{\omega_1^2\omega_2^2-\gamma^2(\omega_1^2+\omega_2^2)-3\gamma^4}{(\omega_1^2+\gamma^2)^2(\omega_2^2+\gamma^2)^2},
\end{equation}
where $\gamma=\Omega_L^2/(2\lambda)$ is the spin relaxation rate in this limit, see Eq.~\eqref{eq:Omega_star}. This expression coincides with the bispectrum of the telegraph noise~\cite{Sinitsyn-Correlators}. Thus when the quantum Zeno effect is strong, the electron spin is in every time moment either $+1/2$ or $-1/2$, as required for the telegraph noise.

\section{Discussion and conclusion}
\label{sec:concl}

Generally, the spin measurement leads to the anisotropic spin relaxation. In order to reach the quantum limit this should be the only source of the spin relaxation, while there are many other additional spin relaxation mechanisms of an electron in a QD. In weak magnetic field, the dominant spin relaxation mechanisms are the hyperfine interaction~\cite{book_Glazov} and Auger trion recombination~\cite{Occupancy-noise}, they lead to the spin relaxation on the time scale from a few nanoseconds to tens of microseconds. In the same time, the characteristic time scale $1/\varkappa$ is usually of the order of $1$~ps, so there is a time gap of three orders, when the spin relaxation can be neglected. The model of the quantum Zeno effect developed in this paper is valid in this case. In particular, for zero magnetic field, the nondemolition spin measurement can be realized and the quantum limit can be reached using the homodyne detection.

To summarize, we have developed a microscopic theory of the quantum Zeno effect for a continuous measurement of a single electron spin noise in a QD micropillar cavity. We showed that in the limit of small population of the trion states, the suppression of the spin precession in magnetic field is described by a single parameter, the measurement strength given by Eq.~\eqref{eq:Zeno_rate}. We demonstrated that the quantum Zeno effect results from the destruction of the entanglement between the electron spin and photon polarizations. We derived the general expression for the optical spin measurement rate analyzing the Stokes parameters of the transmitted light, Eq.~\eqref{eq:Gamma_t}. We demonstrated that the high order spin statistics qualitatively change during the transition from the regime of quantum coherent dynamics to the quantum Zeno phase and agree with the telegraph noise for the strong continuous spin measurement. And finally, we demonstrated that the quantum limit can reached in this system for any probe frequency using the homodyne nondemolition spin measurement.

\begin{acknowledgments}
We gratefully acknowledge the fruitful discussions with A. N. Poddubny and the partial financial by the RF President Grant No. MK-1576.2019.2 and the Foundation for the Advancement of Theoretical Physics and Mathematics ``BASIS''.
The numerical calculations by N.V.L. were supported by the Russian Science Foundation Grant No. 19-12-00051.
The analytical calculations by D.S.S. were supported by the Russian Science Foundation Grant No. 19-72-00081.
\end{acknowledgments}

\renewcommand{\i}{\ifr}
\bibliography{Zeno}


\end{document}